\documentclass[12pt]{iopart}

\usepackage{latexsym}
\usepackage{amssymb}
\usepackage{amsbsy}
\usepackage{mathrsfs}
\usepackage[pdftex]{graphicx}
\usepackage{epstopdf}
\usepackage[english]{babel}

\newcommand{\insa}{Universit\`e de Toulouse,
INSA-CNRS-UPS, LPCNO, 135 avenue de Rangueil,
31077 Toulouse, France}

\newcommand{\uama}{\'Area de F\'isica Te\'orica y
Materia Condensada, Universidad Aut\'onoma Metropolitana
Azcapotzalco, Av. San Pablo 180, Col. Reynosa-Tamaulipas,
02200 Cuidad de M\'exico, M\'exico}

\begin{document}

\title{Machine learning assisted GaAsN circular polarimeter }
\author{A. Aguirre-Perez$^1$, R. S. Joshya$^2$, H. Carr\`ere$^2$,
      X. Marie$^2$, T. Amand$^2$, A. Balocchi$^2$, A. Kunold$^1$}
\address{$^1$\uama\\ $^2$\insa}

\begin{abstract}
We demonstrate the application of a two stage machine learning
algorithm that enables to correlate the electrical signals
from a GaAs$_x$N$_{1-x}$ circular polarimeter with the intensity,
degree of circular polarization and handedness of an incident
light beam.
Specifically, we employ a multimodal logistic regression
to discriminate the handedness of light and
a 6-layer neural network to establish the relationship between
the input voltages,
the intensity and degree of circular polarization.
We have developed a particular neural network training strategy 
that substantially improves the accuracy of the device.
The algorithm was trained and tested on theoretically generated
photoconductivity and on photoluminescence experimental results.
Even for a small training experimental dataset (70 instances),
it is shown that the proposed algorithm
correctly predicts linear, right and left circularly
polarized light misclassifying less than $1.5\%$ of the cases
and attains an accuracy larger than $97\%$ in the vast majority
of the predictions ($92\%$) for intensity and degree of circular polarization.
These numbers are significantly improved for the larger theoretically
generated datasets (4851 instances).
The algorithm is versatile enough that it can be easily adjusted to
other device configurations where a map needs to be established between
the input parameters and the device response.
Training and testing data files as well as the algorithm 
are provided as supplementary material.
\end{abstract}
%\keywords{Circular polarimetry, Spin-dependent recombination, GaAsN}

\maketitle

\section{Introduction}
Circular polarimetry, the determination of the handedness of light,
plays a key role in a wide range of applications.
Owing to the strong interaction of circularly polarized light
with chiral organic entities,
circular dichroism spectroscopy, a technique that requires
the determination of the circular polarization state of light,
has been applied in a remarkably large number of biology-related areas.
This exploratory tool
has been used in biological
imaging\cite{sparks2019classical,khorasaninejad2016multispectral},
medical diagnosis techniques
\cite{whittaker1994quantitative,10.1117/12.2288761,10.1117/1.JBO.21.5.056002},
food technology \cite{yan2021solubility,chen2021new},
virus research \cite{gaurav2021role},
pharmaceutics \cite{savile2010biocatalytic},
environmental sciences\cite{LI2021144858}
and protein secondary structure determination\cite{brishti2021structural}.
Other fields as quantum information technology
\cite{gao2012observation,bhaskar2020experimental,togan2010quantum,PhysRevB.92.081301},
and magnetic material characterization\cite{radisavljevic2017structural,HAW2021158050}
have also largely benefited from the detection and characterization
of circularly polarized light.
In traditional circular polarimetry light has to go through
many optical elements and movable parts\cite{basiri2019nature}
that have hindered the miniaturization and integration of circular polarimeters
to standard electronics.
The need of many stages to detect circularly polarized light,
is mainly due to the rare occurrence of chirality in
standard optoelectronic materials\cite{joshya2021}.
Precisely, recent attempts to build smaller
circular photodetectors  consist
in incorporating chiral structures into a matrix.
Such is the case of
chiral metamaterials \cite{Akbari:18,hu2017all,zhao2012twisted,
basiri2019nature,sobhani2013narrowband,li2015circularly},
chiral organic semiconductor transistors\cite{yang2013circularly},
chiral plasmonic flat devices \cite{Bai:s} and
chiral MoSe$_2$ metasurfaces\cite{C9NR10768A}.
Various other alternatives have been advanced to implement
miniature circular polarimeters. Among others, we have
silicon-on-insulator photodectectors
\cite{DONG2020125598,FullyintegratedCMOScompatiblepolarizationanalyzer,
Lin:19},
perovskite-based photodiode\cite{C7MH00197E,ishii2020direct}
photovoltaic spin-Hall effect polarimeters
\cite{doi:10.1063/1.2199473,doi:10.1063/1.3327809,doi:10.1063/1.4929326}
and Fe/MgO/Ge heterostructures\cite{doi.org/10.1002/adma.201104256}.
Despite having some positive qualities,
these devices suffer either from
low discrimination between left-handed (LHCP)
and right-handed (RHCP) light,
weak quantum efficiency,
low operation currents,
narrow wavelength range,
extremely complicated architectures
or low saturation powers.
In some cases an off-chip detector is required
or the device operates at cryogenic temperatures.
A very thorough comparison between existing
circular polarimeter proposals can be found in
Ref. \cite{PhysRevApplied.15.064040}.

In two previous articles\cite{joshya2021,PhysRevApplied.15.064040}
we have developed and tested the concept of a circular polarization
detection device based on the spin dynamics of Ga$^{2+}$ paramagnetic
centers in GaAs$_{1-x}$N$_x$.
The GaAs$_{1-x}$N$_x$ semiconductor epilayer acts as a
photoconductive device, sensitive to the chirality of light
thanks to two effects: the hyperfine coupling between
bound electrons and nuclei in Ga$^{2+}$ centers and
the spin-dependent filtering of conduction band (CB) electrons through
them. Roughly, the first effect enables to discriminate
the handedness of light and the second endows the device
with sensitivity to the degree of circular polarization (DCP).

The spin-filtering effect is primarily
caused by the spin-dependent recombination (SDR)
\cite{PhysRevB.6.436}
%,WEISBUCH1974141,PhysRevB.30.931,
%Kalevich2005,doi:10.1063/1.2150252,Kalevich2007,
%doi:10.1002/pssa.200673009,Zhao_2009,wang2009room,
%KALEVICH20094929,
%doi:10.1063/1.3186076,doi:10.1063/1.3273393,doi:10.1063/1.3275703,
%doi:10.1063/1.3299015,Ivchenko_2010,PhysRevB.83.165202,
%PhysRevB.85.035205,doi:10.1063/1.4816970,Kalevich2013,
%puttisong2013efficient,PhysRevB.87.125202,PhysRevB.90.115205,
%PhysRevB.91.205202,Ivchenko2016,PhysRevB.95.195204,
%PhysRevB.97.155201,Sandoval-Santana2018,ibarra2018spin,
-\cite{chen2018room}
that CB electrons undergo through Ga$^{2+}$ centers.
The common consensus is that 
through the introduction of a small percentage of N,
Ga$^{2+}$ centers are formed
in some interstitial sites of the GaAs lattice
\cite{doi:10.1063/1.3275703,wang2009room,
doi:10.1063/1.4816970,ibarra2018spin}.
The presence of a bound 4s$^1$ electron in the outer orbital
of the Ga$^{2+}$ inhibits the recombination of CB electrons with
the same spin orientation and, at the same time, eases the recombination
of CB electrons with the opposite spin.
The great contrast between the capture rates of electrons
with opposing spin in the centers promotes
the formation of an excess population of
spin-polarized CB electrons generated by optical orientation.
This mechanism enables to control the CB electron population
and thus the conductivity through the DCP
of an incident beam of light and the subsequent spin polarization of
CB electrons.
Though efficient, this process is independent of the
handedness of the incident circularly polarized light
due to the isotropic nature of the hyperfine interaction
between the bound electron and the nuclei.
This mechanism is therefore only capable of providing information
on the DCP of light.
However, the symmetry of the
hyperfine interaction can be broken by a moderate magnetic field
in Faraday configuration generating a chiral photocurrent.

So far we have theoretically and experimentally demonstrated
that the chiral photoconductivity (PC) is both
sensitive to the handedness
and the DCP
of the incident light\cite{joshya2021}.
In order to fully implement a circular polarimeter
based on the spin-filtering effect it is essential
to provide a method that translates the photocurrent information
into handedness, degree of polarization and intensity of the
incident light.
This goal can be achieved through artificial neural network (NN)
machine learning techniques\cite{aggarwal2018neural,nielsen2015neural}.

In the past few years, deep learning, a branch of artificial intelligence,
has remarkably developed partly owing to the faster computer capabilities
and to the availability of vast amounts of digital data produced in the web.
In contrast to the conventional approach to programming,
in which a complex task is broken up into many small ones,
precisely defined and easy to code,
NNs simulate the learning mechanisms in biological organisms.
During the learning process, the NN parameters are adjusted,
giving it the capacity of reproducing all the possible outputs
from the corresponding inputs provided by large datasets.

These techniques have been applied
to the solution of a wide range of problems in
optoelectronics as
the identification of light sources\cite{doi:10.1063/1.5133846},
the detection and classification of defects in transparent substrates
\cite{Chang_Chien_2019},
the calibration of single-photon detectors
\cite{PhysRevApplied.15.044003},
the characterization and inverse design of photonic crystals
\cite{Christensen4192,SoBadloeNohBravoAbadRho1057}
and
the detection enhancement of quadrant photodiodes
\cite{CAO2021165971}.
More specifically, machine learning has also proven to be quite useful
in the context of circular polarization detection in
improving the performance and accuracy of a Stokes polarimeters
\cite{doi:10.1021/acsphotonics.8b00295},
analyzing the near-field intensity distribution to determine
the arbitrary states of polarization in a circular polarimeter
based on plasmonic spirals\cite{Zhou:19}
and
in investigating the circular dichroism properties in the higher-order
diffracted patterns of two-dimensional chiral metamaterials
\cite{DuYouZhangTaoHaoTangZhengJiang+2021+1155+1168}.

In this paper we demonstrate the application of a neural-network-based
two-stage machine learning algorithm
that in a single shot translates the PC input
into handedness, degree of polarization and intensity data in
a GaAs$_x$N$_{1-x}$ device. The first stage of the algorithm
consists of a multimodal logistic regression, a probabilistic perceptron model,
that discriminates the handedness of light: RHCP, LHCP or linear polarization.
The second stage is a 6-layer NN that maps between the PC
input data and the output data corresponding to circular polarization degree and intensity.
A specially customized training strategy was used to correctly characterize
the NN parameters.
The algorithm was tested on experimental and theoretically generated datasets.

The paper is organized as follows.
In Sec. \ref{section:plandpc}
we outline how the intensity and the degree of circular
polarization affect the PC
and the PL as a function of the
applied magnetic field. We also describe the
main features of the intensity and degree of circular polarization
isolines. The machine learning algorithm is sketched in
Sec. \ref{section:algorithm}.
More specifically, the logistic regression
and the NN are presented in
Secs. \ref{section:logreg}
and \ref{section:neuralnetwork}, respectively.

\section{Photoluminescence and photoconductivity}
\label{section:plandpc}

\begin{figure}
\flushright
\includegraphics[width=0.65\textwidth]{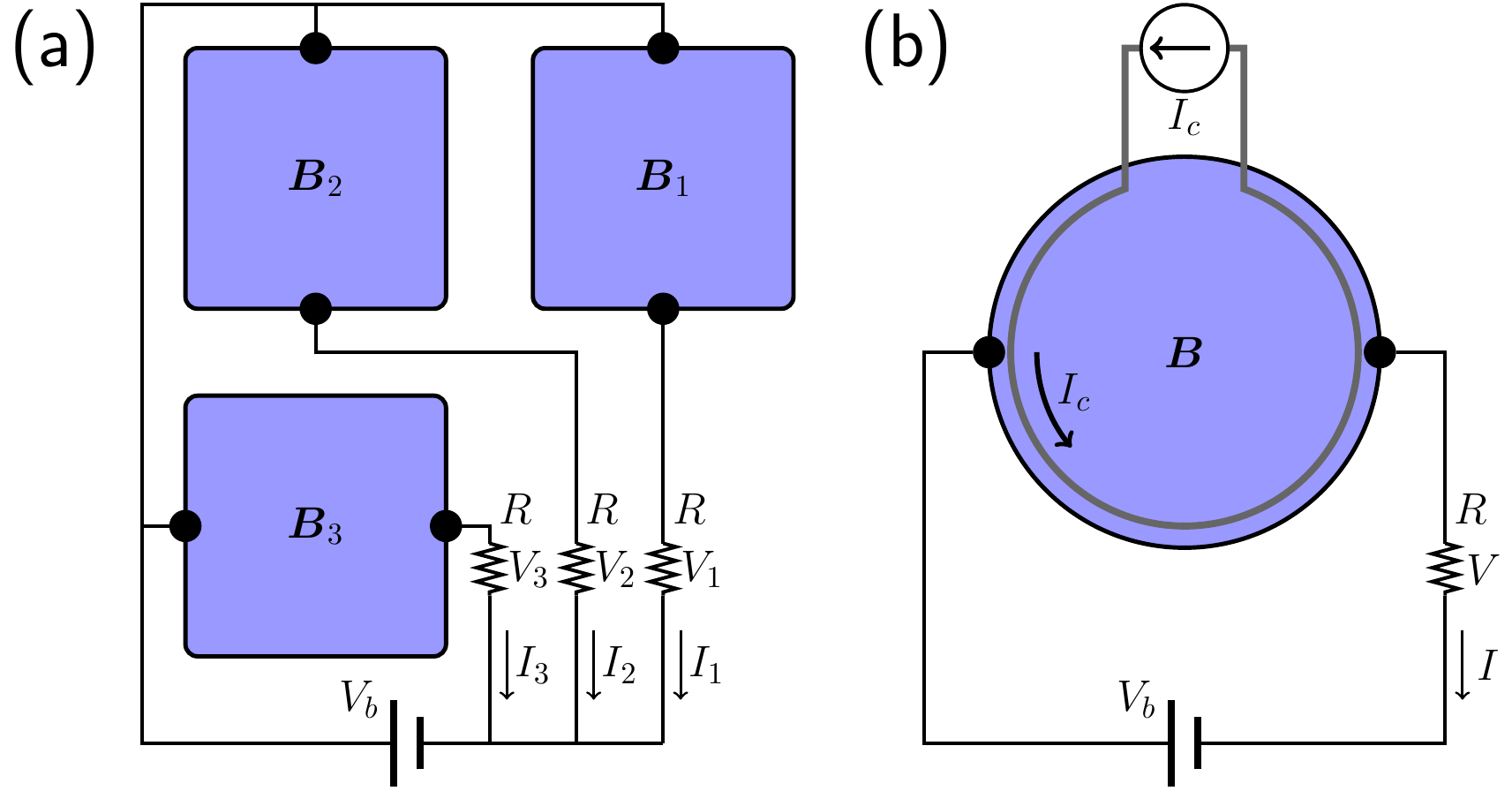}
\caption{Two possible configurations of the circular polarimeter.
(a) Three separate GaAs$_{1-x}$N$_x$ slabs are subject
to the magnetic fields $B_1=-50mT$, $B_2=50mT$, $B_3=150mT$
through three different permanent magnets.
(b) The variable magnetic field produced by the
micro-coil is experienced by a single
GaAs$_{1-x}$N$_x$ slab.}
\label{figure1}
\end{figure}

We have experimentally\cite{joshya2021}
and theoretically\cite{PhysRevApplied.15.064040}
demonstrated
that GaAs$_x$N$_{1-x}$ exhibits sensitivity to
different features of light depending on
the intensity of the external magnetic field $B$
it is subjected to.
According to this principle, the two different
device configurations depicted in Fig. \ref{figure1}
have been proposed.
In the first, Fig. \ref{figure1} (a),
the intensity and the degree of circular polarization
are determined from the conductivities
of three  GaAs$_x$N$_{1-x}$
slabs, each one subject to the magnetic
field of a different permanent magnet\cite{PhysRevApplied.15.064040}.
In the second configuration, Fig. \ref{figure1} (b),
a variable external magnetic field is provided
by a micro-coil
built on a single GaAs$_x$N$_{1-x}$
slab\cite{joshya2021}.

In the absence of a magnetic field  the PL
and the PC of GaAs$_x$N$_{1-x}$
are mostly sensitive to the illumination power and the
degree of circular polarization.
Under a moderate magnetic field ($50$\, mT), the combined
effects of SDR, Zeeman interaction and hyperfine coupling
occurring in $Ga^{2+}$ centers along with the dipole-dipole interaction
between neighbouring Ga atoms make GaAs$_x$N$_{1-x}$
very responsive to the handedness of light.
At higher yet moderate magnetic fields
($>100$\, mT) the PL and the PC are
also strongly influenced by the incidence angle of the light source.

The photoluminescence (PL) is determined from
\begin{equation}
    P_L=c_rnp,
    \label{eq:photolum}
\end{equation}
where $c_r=1/N_0\tau_r$ is the band to band recombination
coefficient, $N_0$ is the total number of traps in the
sample and $\tau_r$ is the electron-hole recombination time
at low power.
The electron and hole densities $n$ and $p$,
are obtained from the numerical solution
of a master equation model for the density
matrix developed by us\cite{PhysRevApplied.15.064040}.
Within the framework of Drude theory the PC
is given by
\begin{equation}
  P_{C}
      \approx e\left[n\mu_e
      +p\left(\mu_{lh}+\mu_{hh}\right)/2\right],
       \label{eq:drude01}
\end{equation}
where $e$ is the electron charge,
$\mu_e=300$\,cm$^2/$Vs,
and $\mu_{hh}=50$\,cm$^2/$Vs \,\,
\cite{doi:10.1063/1.2798629,Dhar_2007,
doi:10.1063/1.2927387,5411156,
https://doi.org/10.1002/pssc.201200383,Patan_2009}
are the electron, light hole and heavy hole mobilities.
In contrast to Eq. (\ref{eq:drude01}), the exact Drude expression
for the longitudinal conductivity
contain magnetic field-dependent terms.
However, for low magnetic fields
($\approx 150$\,mT) $\mu_eB$, $\mu_{hh}B$ and
$\mu_{hh}B \ll 1$, rendering magnetic field-dependent terms negligible.
The magnetic field dependence, though, enters  $P_C$
through the electron and hole densities $n$ and $p$
due to the modulations in the spin-filtering
efficiency induced
by the interplay between the Zeeman interaction and
the hyperfine coupling experienced by the bound electron
in Ga$^{2+}$ centers.
It is important to note that $P_L$ and $P_C$ follow similar trends,
despite having quite different algebraic expressions
[see Eqs. (\ref{eq:photolum}) and (\ref{eq:drude01})].
This is because the inverted Lorentzian shape
of $P_L$ and $P_C$ as functions of the magnetic field
is also shared by the electron
and hole populations $n$ and $p$.

In the master equation model for the density
matrix, $n$ and $p$ are worked out
directly from the refracted beam inside the sample.
To connect the intensity and degree of circular
polarization of the refracted beam with the
the incident one, we renormalize $P_{\mathrm{exc}}$ and $P_e$
from the incoming ray
according to the Fresnel equations.
More precisely, the excitation power $\tilde{P}_{\mathrm{exc}}$
and the degree of circular polarization $\tilde{P}_e$ experienced
by the GaAs$_x$N$_{1-x}$ sample
are proportional to $P_{\mathrm{exc}}$ and $P_e$ according to
\begin{eqnarray}
    \tilde{P}_{\mathrm{exc}} &=& \frac{n_0}{2}
\left(C_\perp^2+C_\parallel^2\right)P_{\mathrm{exc}},\\
    \tilde{P}_{e} &=&
    \frac{2 C_\perp C_\parallel}{C_\perp^2+C_\parallel^2}P_{e},
\end{eqnarray}
where $n_0$ is the refractive index of GaAs$_x$N$_{1-x}$.
The coefficients $C_\perp$ and $C_\parallel$ are the
relative amplitudes of the refracted ray's electric field
with respect to the amplitude of the incoming ray's electric field
perpendicular and parallel to the incidence plane, respectively.
These can be worked out from the Fresnel equations. Neglecting
their imaginary parts, they are given by
\begin{eqnarray}
C_\perp &=& \frac{2 \cos\theta}
{\cos\theta +\sqrt{n_0^2-\sin^2\theta}}\, ,\\
C_\parallel &=& \frac{2 n_0\cos\theta}
{n_0^2\cos\theta +\sqrt{n_0^2-\sin^2\theta}}\, ,
\end{eqnarray}
where $\theta$ is the incidence angle.
The direction of the incidence beam
also modifies $P_L$ and $P_C$
as the refraction angle $\theta_r$ between the
refracted ray and the external magnetic field
has a very significant influence on the efficiency
of the spin-filtering effect.
Under normal incidence, where the magnetic field and
and the photogenerated spin are parallel (Faraday configuration),
the SDR is governed by the amplification of the spin-filtering 
effect \cite{PhysRevB.85.035205}.
In contrast, when the magnetic field and the incidence
line are perpendicular (Voigt configuration),
the Hanle effect is dominant \cite{KALEVICH20094929}.
Both effects yield drastically different $P_L$ and
$P_C$ features; the Hanle effect is characterized by
a Lorentzian curve centered in $B=0$,
whereas the amplification of
the spin-filtering effect results in an inverted
Lorentzian function centered in $B\approx \pm 50mT$.
For oblique angles, a combination of both
effects shapes $P_L$ and $P_C$ as
functions of $B$\cite{Ivchenko2016}.
The refraction angle $\theta_r$ can readily be obtained
from Snell's law as $\theta_r=\arcsin(\sin(\theta)/n_0)$.

\begin{figure}
\flushright
\begin{minipage}{0.25\textwidth}
\includegraphics[width=0.75\textwidth]{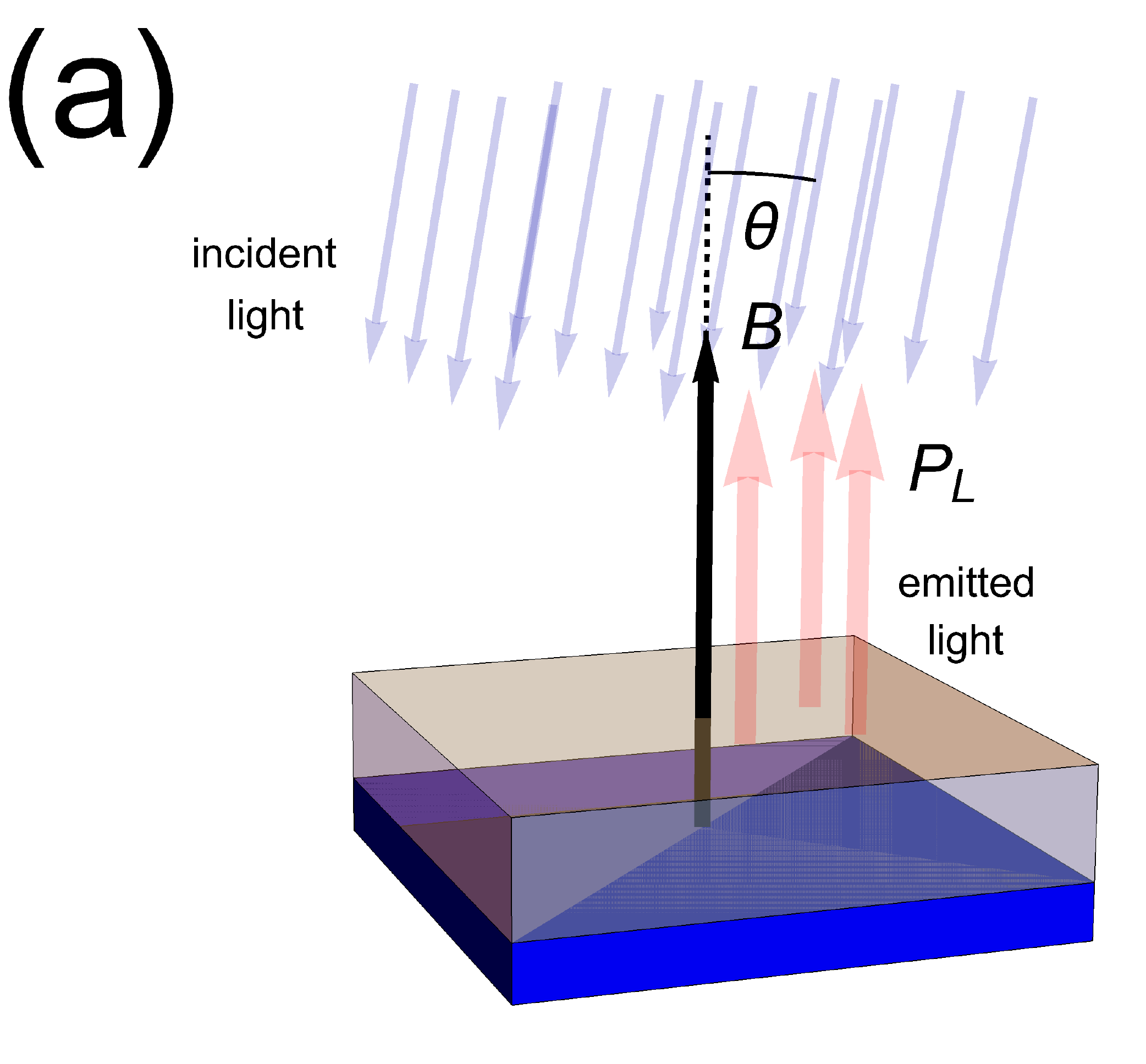}\\
\vspace{0.2cm}
\includegraphics[width=1.0\textwidth]{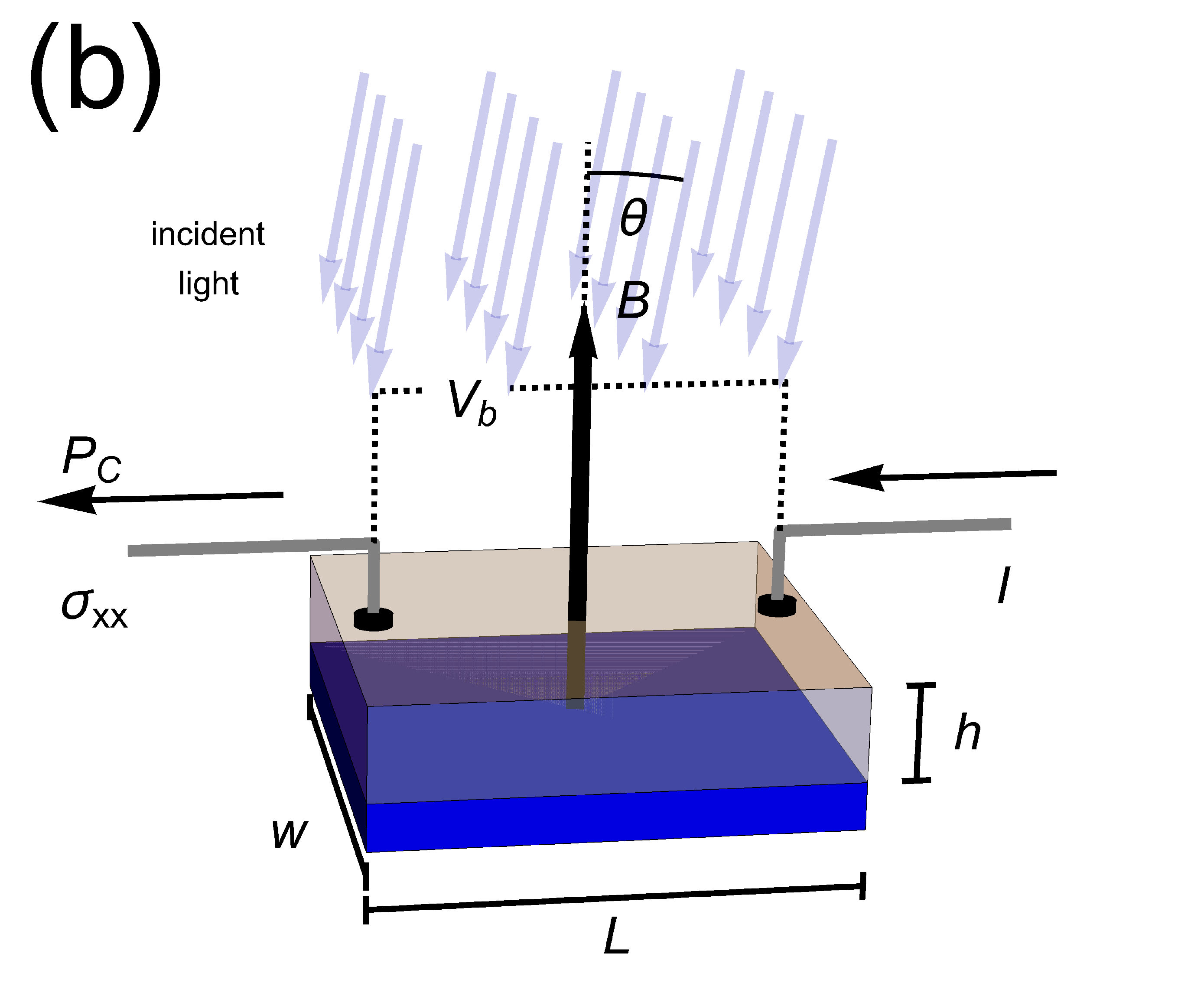}
\end{minipage}
\begin{minipage}{0.65\textwidth}
\includegraphics[width=1.0\textwidth]{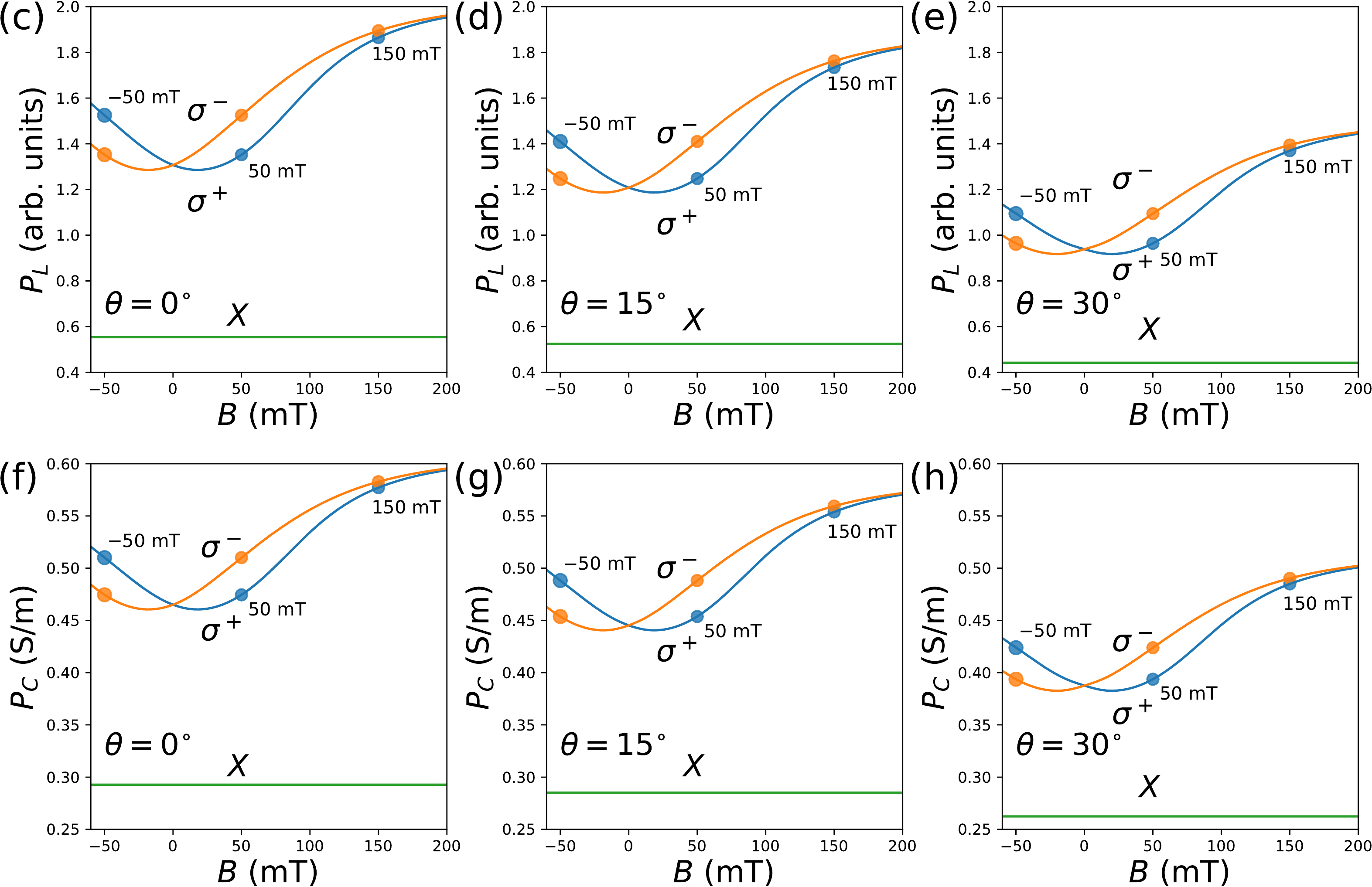}
\end{minipage}
\caption{Theoretical results of the
photoluminescence and photoconductivity as functions
of the external magnetic field.
Sketches of the (a) photoluminescence
and (b) photoconductivity experimental setups.
Photoluminescence
as a function of the external magnetic field $B$ at incidence angles
(c) $\theta=0^\circ$, (d) $\theta=15^\circ$ and (e) $\theta=30^\circ$.
Photoconductivity
as a function of the external magnetic field $B$ at incidence angles
(f) $\theta=0^\circ$, (g) $\theta=15^\circ$ and (h) $\theta=30^\circ$.
All the plots correspond to an incident light beam of
$20$\,mW and a degree of circular polarization of $80\%$ for
right ($\sigma^+$), left ($\sigma^-$) circularly polarized light
and linearly ($X$) polarized light. The points
at $-50$\, mT, $50$\, mT and $150$\, mT are indicated in the plots.}
\label{figure2}
\end{figure}

Figure \ref{figure2} presents the theoretical
results for
the PL and the PC
as functions of the external magnetic field
for right and left circularly polarized sources 
with fixed intensity and degree of
circular polarization.
Sketches of the corresponding experimental setups
for the PL and PC
are exhibited in panels (a) and (b).
The results for different incidence angles
are shown for $P_L$ in panels (c-e) and for $P_C$ in (f-g).
The points for $P_L$ and $P_C$ at $-50$\,mT, $50$\,mT,
and $150$\,mT are marked in panels (c-h).
The first thing to note is that for $\sigma^+$
the difference $P_C(-50\,\mathrm{mT})-P_C(50\,\mathrm{mT})$ is positive
while for $\sigma^-$ it is negative
and, therefore, it is
sensitive to the handedness of light.
The same behaviour is observed for $P_L$.
In the sections to follow,
this is used as the key principle
for the logistic regression to identify
the handedness of light.

\begin{figure}
\flushright
\includegraphics[width=0.9\textwidth]{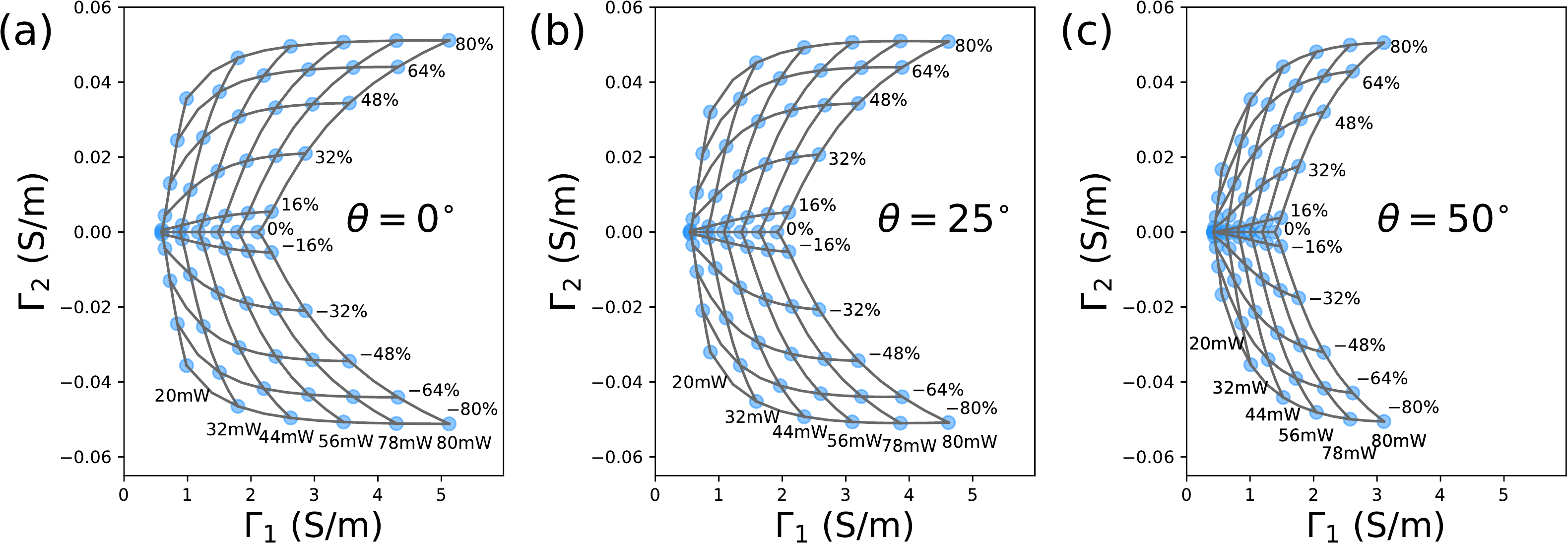}
\caption{Theoretical calculations
of the intensity and degree of circular polarization isolines
as a function of $\Gamma_1$ and $\Gamma_2$ for
incidence angles (a) $\theta=0^\circ$, (b) $\theta=25^\circ$ and
(c) $\theta=50^\circ$.}
\label{figure3}
\end{figure}

Three parameters are needed in order to gain sensitivity
to the intensity and the degree of circular polarization
regardless of the incidence angle of the light stimulation.
A very convenient set of parameters
based on the PC
is given by
\begin{eqnarray}
\Gamma_1 &=& P_C(-50\,\mathrm{mT})+P_C(50\,\mathrm{mT}).
    \label{eq:gamma1} \\
\Gamma_2 &=& P_C(-50\,\mathrm{mT})-P_C(50\,\mathrm{mT}),\\
\Gamma_3 &=& P_C(150\,\mathrm{mT})/P_C(50\,\mathrm{mT}).
\label{eq:gamma3}
\end{eqnarray}
and similarly for PL measurements by
\begin{eqnarray}
\Gamma_1 &=& P_L(-50\,\mathrm{mT})+P_L(50\,\mathrm{mT}).
    \label{eq:plgamma1} \\
\Gamma_2 &=& P_L(-50\,\mathrm{mT})-P_L(50\,\mathrm{mT}),\\
\Gamma_3 &=& P_L(150\,\mathrm{mT})/P_L(50\,\mathrm{mT}).
\label{eq:plgamma3}
\end{eqnarray}

\begin{figure}
\flushright
\includegraphics[width=0.9\textwidth]{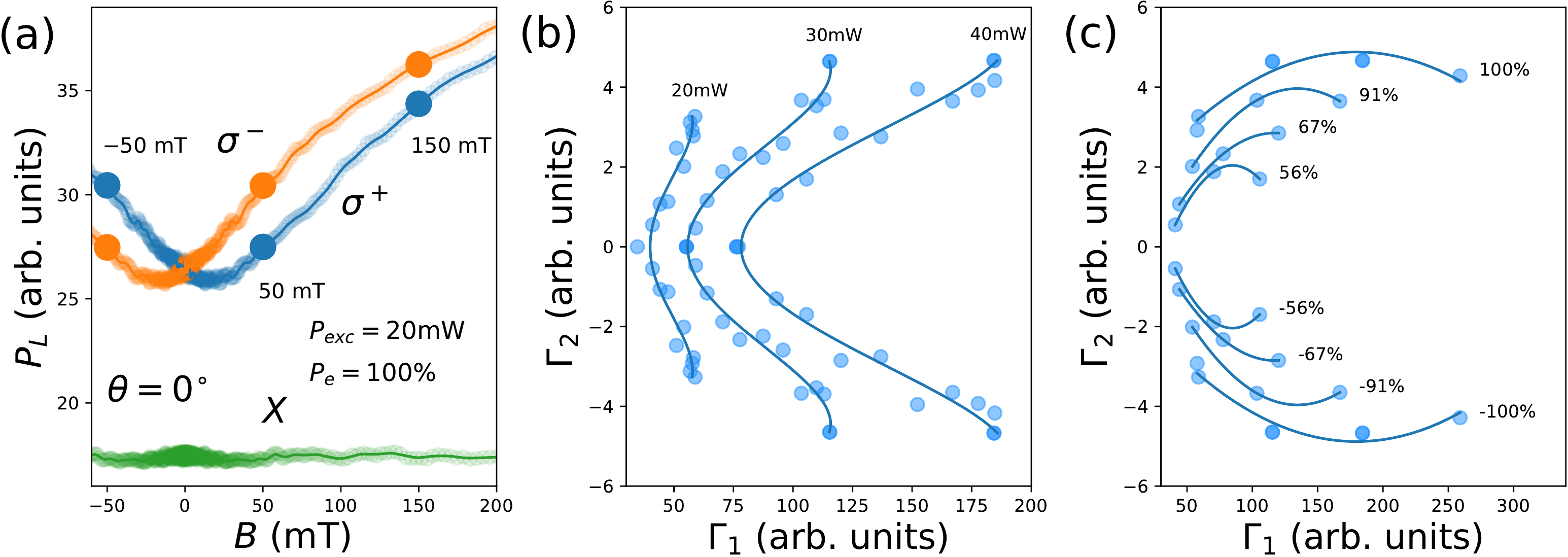}
\caption{Experimental results for the
(a) photoluminescence as functions
of the external magnetic field.
Results for linearly (green),
right (blue) and
left circularly polarized light sources are exhibited.
(b) Intensity and (c) degree of circular polarization isolines
as a function of $\Gamma_1$ and $\Gamma_2$ at normal incidence.
}
\label{figure4}
\end{figure}

By extracting the PCs at $-50$\, mT and $50$\, mT
at different intensities and degrees of circular polarizations
(as shown in Fig. \ref{figure2})
we obtain the power and degree of circular polarization
isolines as functions of $\Gamma_1$ and $\Gamma_2$
shown in Fig. \ref{figure3}.
Similar features are displayed by the
experimental data for the PL in
Fig. \ref{figure4}.
Panel (a) shows $P_L$ as a function
of the external magnetic field
under a light stimulation of $20$\,mW at normal incidence.
Results for linearly (green),
right (blue) and
left (ornage) circularly polarized light sources are shown here.
Panels (b) and (c) of Fig. \ref{figure4}
present the intensity and degree of polarization
isolines in the $\Gamma_1$ and $\Gamma_2$ parameter
space extracted from $P_L(B)$ curves at diverse
powers and degrees of circular polarization.
We observe in the experimental results, as well
as in the theoretical ones,
that $\Gamma_1$ and $\Gamma_2$ predominantly
parametrize intensity and degree of circular polarization
provided that the incidence angle remains fixed.
However, when the incidence angle is varied, as the
theoretical plots show in Fig. \ref{figure3}, the
isolines are strongly distorted.
An extra parameter is thus needed to incorporate
information on the incidence angle.
The parameter $\Gamma_3$
is thus taken in the large magnetic field region ($150$\, mT)
where the $P_C(B)$ curve is most sensitive to
the incidence angle
as it shifts from
an upward to a downward Lorentzian
for the above-mentioned reasons.

In principle one could attempt to
determine the whole set of
$P_e$ and $P_{\mathrm{exc}}$
isolines for each possible value of $\Gamma_3$.
Fortunately, the use of a NN
spares us the work of determining
the precise dependence of the power and the
degree of circular polarization in terms of
$\Gamma_1$, $\Gamma_2$ and $\Gamma_3$.
It is sufficient
that the points are scattered enough in the parameter space
and no precise knowledge of the shape of the isolines
is needed.

\begin{figure}
\flushright
\includegraphics[width=0.9\textwidth]{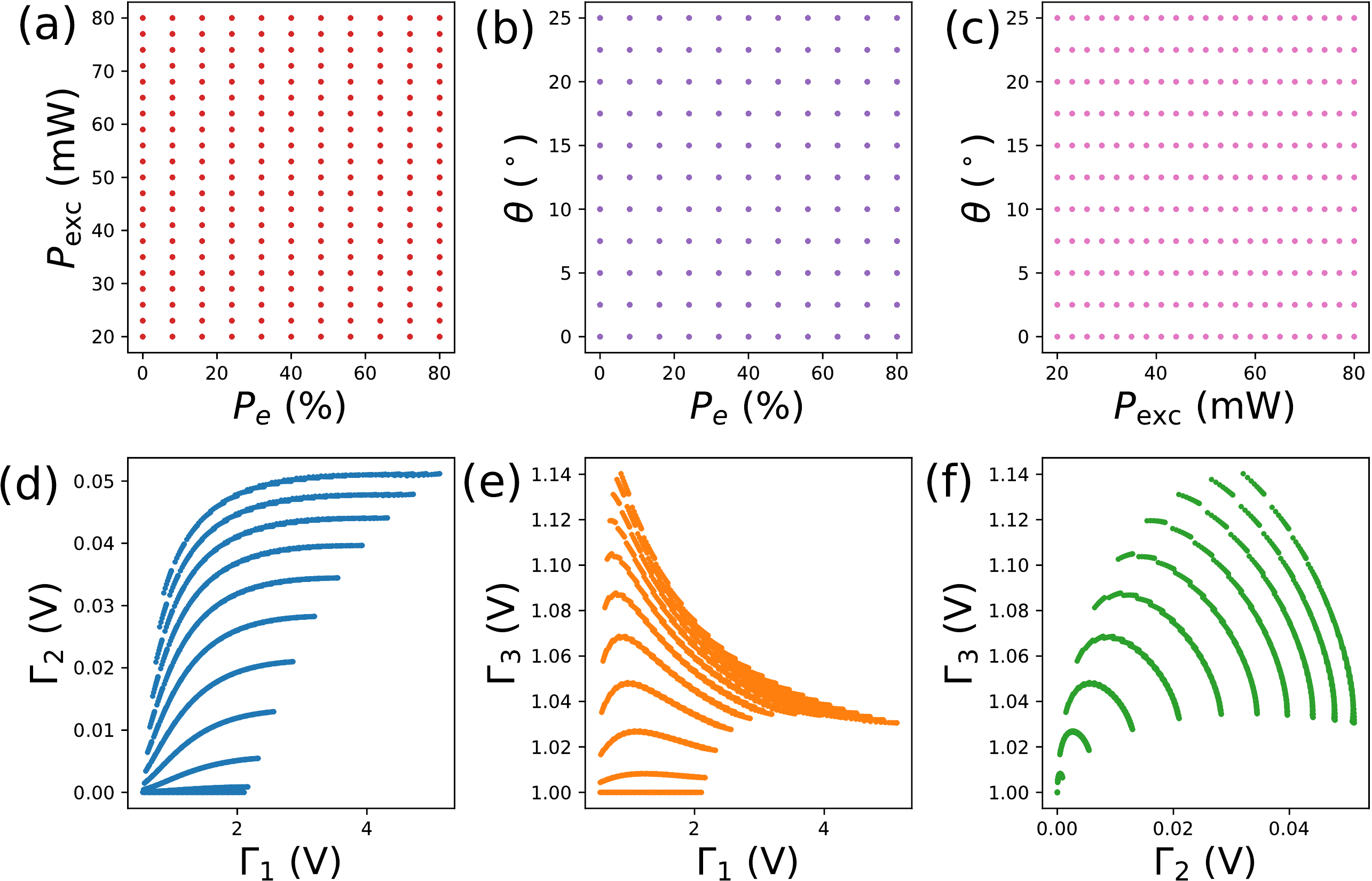}
\caption{Map of the degree of circular polarization $P_e$,
intensity $P_{\mathrm{exc}}$ and incidence angle $\theta$
to the parameter space $\Gamma_1$ $-$ $\Gamma_2$ $-$ $\Gamma_3$.
}
\label{figure5}
\end{figure}

Figure \ref{figure5} shows how $P_e$, $P_{\mathrm{exc}}$
and $\theta$ are mapped into $\Gamma_1$, $\Gamma_2$ and $\Gamma_3$.
The regular grid of ordered triads of power, incidence angle
and degree of circular polarization [panels (a-c)]
are mapped into ordered triads of $\Gamma_1$, $\Gamma_2$ and
$\Gamma_3$ [panels (d-f)].
Notwithstanding the complexity of the patterns formed
by the dots in the
$\Gamma_1$ $-$ $\Gamma_2$ $-$ $\Gamma_3$
parameter space, the proposed NN is
able to establish the correspondence 
to the $P_e$, $P_{\mathrm{exp}}$ ordered pairs
as we show in the following sections.

\section{Machine learning algorithm.}
\label{section:algorithm}
In this section we propose a two stage algorithm that
determines the handedness, degree of circular polarization and intensity
of an incident light beam.
Roughly, the first stage determines the handedness and the
second calculates the degree of circular polarization and intensity
of the light stimulation.

The first stage,
a multimodal logistic regression, classifies 
the $\Gamma_1$, $\Gamma_2$, $\Gamma_3$ triads
according to the handedness of light: $\sigma^+$, $\sigma^-$ or $X$.
Ideally the sign of $\Gamma_2$ would suffice
to determine the handedness of light since,
as we saw in Fig. \ref{figure3},
if $\Gamma_2=0$ light is linearly polarized,
if $\Gamma_2>0$ it is right circularly polarized
and if $\Gamma_2<0$ it is left circularly polarized.
However, experimental uncertainty widens
the fringe for linearly polarized light
and thus, light is linearly polarized if
$\left\vert\Gamma_2\right\vert<\Delta$ and it is circularly
polarized if $\left\vert\Gamma_2\right\vert\ge\Delta$.
Moreover, the width of the fringe $2\Delta$ might depend on
the remaining parameters.
The multimodal logistic regression solves this problem by
automatically establishing the $\Gamma_1$, $\Gamma_2$, $\Gamma_3$
parameter regions that correspond to each handedness.

The second stage consists of
a NN that calculates the degree of polarization and
intensity from the entered $\Gamma_1$, $\Gamma_2$, $\Gamma_3$
parameters.

In general each stage undergoes training and testing phases
through the training and testing datasets.
These are datasets where each
instance is formed of an input and the corresponding
output parameters.

During the training phase the model parameters
are optimized in order to minimize a loss function
typically
through a gradient descent-based algorithm.
The loss function is usually some sort of distance
between the predictions of the model 
with the outputs of the training dataset.
The effectiveness of the trained model
is later evaluated in
the testing phase by contrasting its predictions
with the outputs
of the testing dataset.

To test the predicting capabilities
of the logistic regression and the NN
we have used theoretically and experimentally generated datasets.
The theoretical dataset consists of
$4851$ PC related instances
determined
from the aforementioned density matrix master equation.
It was calculated for a range
of incidence angles that go from normal incidence up to $25^{\circ}$.
Each instance is composed of six numbers: $\Gamma_1$, $\Gamma_2$, $\Gamma_3$, $P_e$,
$P_{\mathrm{exc}}$, $\theta$ and $h$
is where
$h$ is a discrete variable that tags the handedness of light
($h=1$ for $X$, $h=2$ for $\sigma^+$ and $h=3$ for $\sigma^-$).

The parameters $\Gamma_1$, $\Gamma_2$ and $\Gamma_3$ are associated
to the PC through Eqs. (\ref{eq:gamma1})-(\ref{eq:gamma3}).
The experimental dataset is made up of $70$ instances
each containing the same information as the theoretical
one but where $\Gamma_1$, $\Gamma_2$ and $\Gamma_3$
are related to the PL through
Eqs. (\ref{eq:plgamma1})-(\ref{eq:plgamma3}).
Although the experimental dataset is restricted to normal
incidence, it is successfully treated with the same models as the
theoretical one.
The experimental dataset was extracted from PL
measurements as a function of the applied magnetic field.
The sample studied consists of a $100$ nm thick
GaAs$_{1-x}$N$_x$ epilayer ($x=0.021$)  grown by molecular beam epitaxy on a (001)
semi-insulating GaAs substrate and
capped with $10$ nm GaAs.
The excitation light was provided by a $852$ nm laser diode focused
to a $\approx$ 100 $\mu$m diameter spot (FWHM).
The circular polarization degree
of the incident laser light  was set by 
adjusting the angle of a quarter wave plate placed after a
Glan-Taylor polarizer.
A permanent Neodymium magnet
mounted onto a linear stage has been used to apply a magnetic field of
controlled strength onto the sample surface by changing the 
magnet's distance from the sample.
The laser intensity was modulated by a mechanical chopper at 170 Hz.
The PL intensity was measured by
recording the total
intensity,
filtered of the laser scattered light and substrate
contribution using a series of long-pass optical filters, and integrated by 
an InGaAs photodiode.
All the experiments were performed at room temperature.

\subsection{Handedness discrimination through multimodal
logistic regression.}
\label{section:logreg}

The multimodal logistic regression is a probabilistic model that
classifies the instances in terms of
probabilities\cite{aggarwal2018neural}.
In the training and testing datasets
an instance is formed of
$(\Gamma_1,\Gamma_2, \Gamma_3, h)$.

\begin{figure}
\flushright
\includegraphics[width=0.65\textwidth]{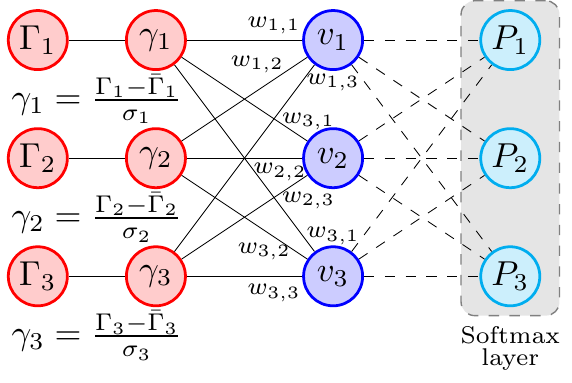}
\caption{Diagram of the multinomial logistic regression used
to detect handedness of light. The dashed lines
in the last layers represent
the probabilities calculated through the softmax function.}
\label{figure6}
\end{figure}

The neural architecture of the multimodal logistic regression
is illustrated in Fig. \ref{figure6}.
In the first layer the
$\Gamma_1$,$\Gamma_2$ and $\Gamma_3$
parameters are entered into the model.
The next layer represents the standardization
of the $\Gamma_1$,$\Gamma_2$ and $\Gamma_3$
parameters by
\begin{equation}
    \gamma_i = \frac{\Gamma_i-\bar{\Gamma}_i}{\sigma_i},
\end{equation}
where $\bar{\Gamma}_i$ and $\sigma_i$ ($i=1,2,3$) are the mean
and the standard deviation of $\Gamma_i$
over all the instances of the training dataset.
Even though it is not essential,
standardization helps reduce the number of steps
necessary to optimize the cost function through the gradient descent
algorithm\cite{raschka2015python}.
This is not an actual neural layer but
including this step in the diagram of the multimodal
logistic regression helps visualize the entire
training procedure.
The activation values $v_i$ of the next layer are
given by
\begin{equation}
    v_i=\sum_{j=1}^3 w_{i,j}\gamma_j+b_i,
    \label{eq:activationvi}
\end{equation}
where $w_{i,j}$ is the weight matrix and $b_i$ is called the bias.
After the optimization process, Eq. (\ref{eq:activationvi})
can be understood as the equations of the hyperplanes that
best separate the parameter regions associated to each handedness.
A dataset is called linearly separable if one can successfully place
each category between two hyperplanes.
This is not the case for the handedness of light in the
$\Gamma_1$,$\Gamma_2$, $\Gamma_3$ parameter space.
One way to overcome this difficulty is to
attach a softmax activation layer at the end of the
NN.
In this approach the model predicts the membership of an instance
in terms of the probabilities given by the softmax function
\begin{equation}
    P_i=\frac{\exp(v_i)}{\sum_{j=1}^3 \exp(v_j)}.
\end{equation}
The cross-entropy,
the negative logarithm of the probabilities,
is used as the cost function in the optimization process.

\begin{figure}
\flushright
\includegraphics[width=0.65\textwidth]{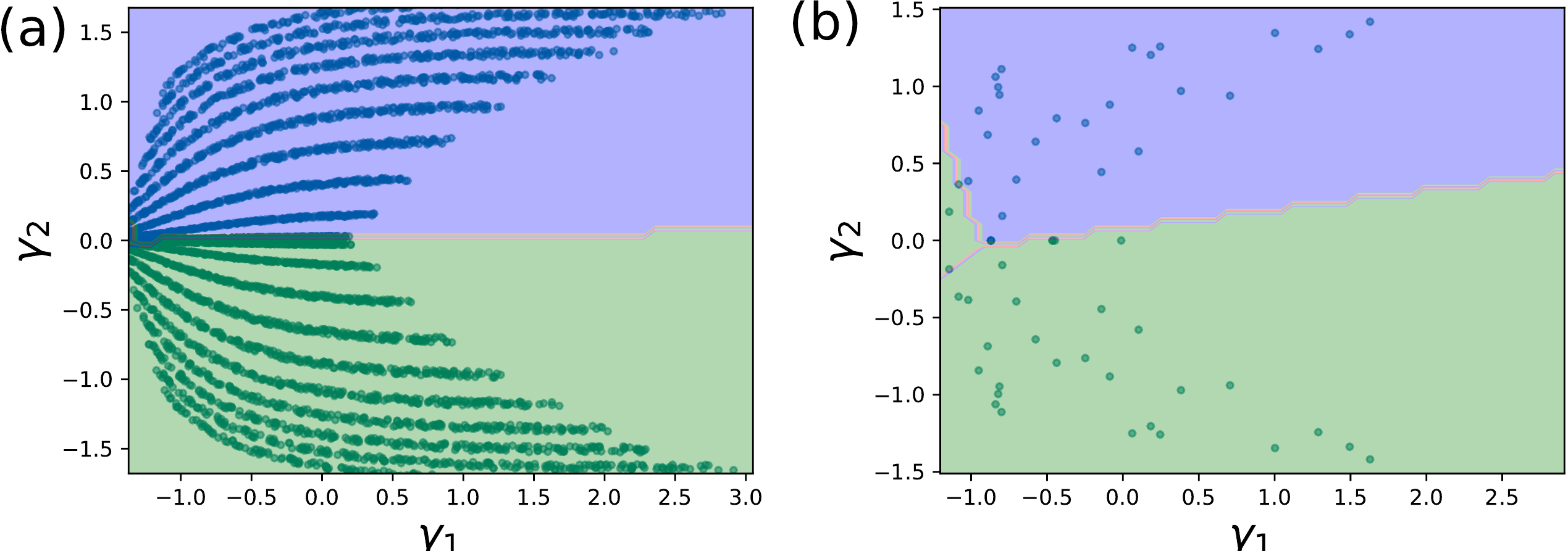}
\caption{Decision regions of handedness for the
(a) theoretical and (b) experimental datasets
as functions of the standardized $\gamma_1$ 
and $\gamma_2$.
The right and left circularly polarized light regions are
shaded in blue and green respectively. The narrow
red fringe corresponds to the linearly polarized region.
The dots correspond to the (a) theoretical and (b) experimental
test datasets.}
\label{figure7}
\end{figure}

\begin{figure}
\flushright
\includegraphics[width=0.65\textwidth]{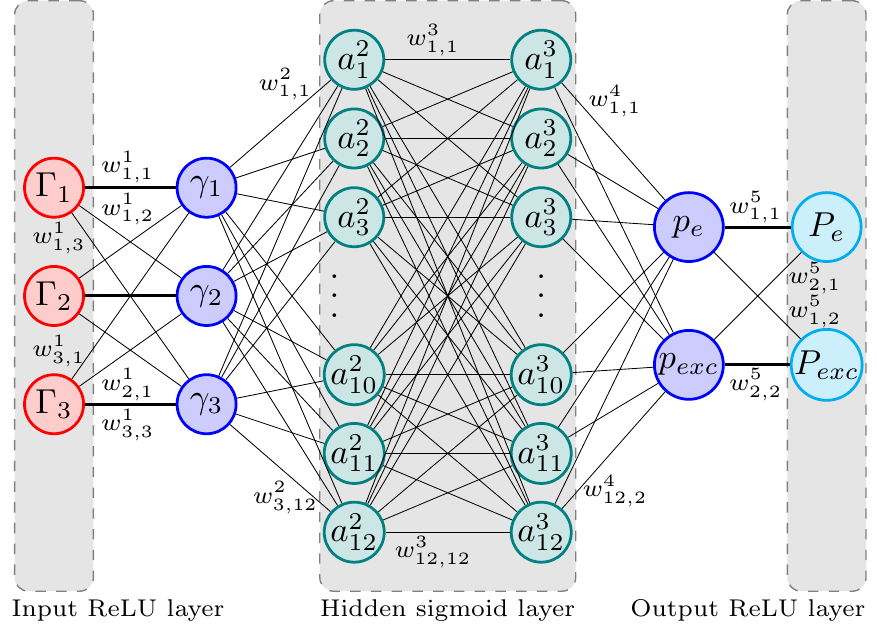}
\caption{Diagram of the neural network.}
\label{figure8}
\end{figure}

Figure \ref{figure7} maps the three decision regions
for each predicted polarization state for the theoretical
[Fig. \ref{figure7} (a)]
and experimental [Fig. \ref{figure7} (b)] datasets.
The points corresponding to right and left circularly polarized light are
located in the upper ($\gamma_2>0$) and lower ($\gamma_2<0$)
portions of the plot respectively. The theoretical points associated
to linearly polarized light fall inside a very narrow
fringe close to $\gamma_2=0$. Due to experimental error
the linearly polarization fringe deviates from $\gamma_2=0$
where no experimental training points are provided.
In the supplementary material we provide
the {\it jupyter notebook} that contains
the code using the {\it sklearn} libraries and
the theoretical and experimental training and testing datasets.
As it can be seen in the algorithm file, the logistic regression
successfully classifies $99.9\%$ of the
points in the theoretical testing dataset and $98\%$ of the points
of the experimental testing dataset. It is worthwhile noting
that despite the small number of instances in the
experimental dataset, the logistic regression is able
to correctly place the points in most of the cases.

\subsection{
Intensity and degree of circular polarization determination
through a neural network.}
\label{section:neuralnetwork}

Determining the intensity and degree of polarization
is easier and simpler if the task is divided
in two for right and left circularly polarized light.
The case of linearly polarized light can be dealt with
as a separate case or together with either handedness of light.
In this approach we bundle linearly polarized light
and right circularly polarized light
and treat the case of left circularly polarized light
separately.
This is very easily achieved since
at this point
the handedness of light has already been established
through the logistic regression.

The NN used to extract the
intensity $P_{\mathrm{exc}}$
and degree of circular polarization
from $\Gamma_1$, $\Gamma_2$ and $\Gamma_3$
is sketched in Fig. \ref{figure8}.
This model is used for either handedness of light.
The first and the last layers (layers 1 and 6) serve
to standardize the input ($\Gamma_1$, $\Gamma_2$ and $\Gamma_3$)
and output ($P_e$ and $P_{\mathrm{exc}}$) data.
The hidden layers (layers 2-5) do the actual work of
transforming the input into an intensity and degree of polarization
output.

\begin{figure}
\flushright
\includegraphics[width=0.65\textwidth]{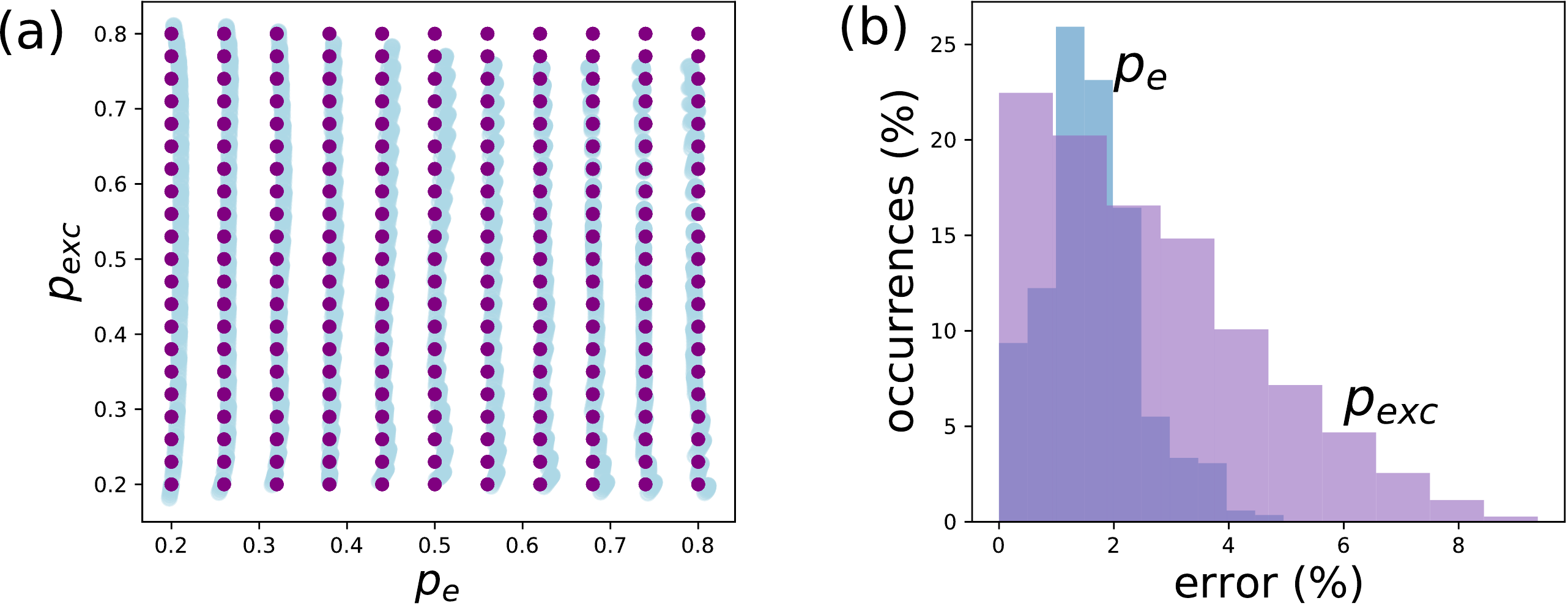}
\caption{Linearly polarized and right circularly polarized light
(a) predictions and (b) error distribution of
the theoretical testing dataset for the training of the
sub-neural network comprised of layers 2-5.
}
\label{figure9}
\end{figure}

\begin{figure}
\flushright
\includegraphics[width=0.65\textwidth]{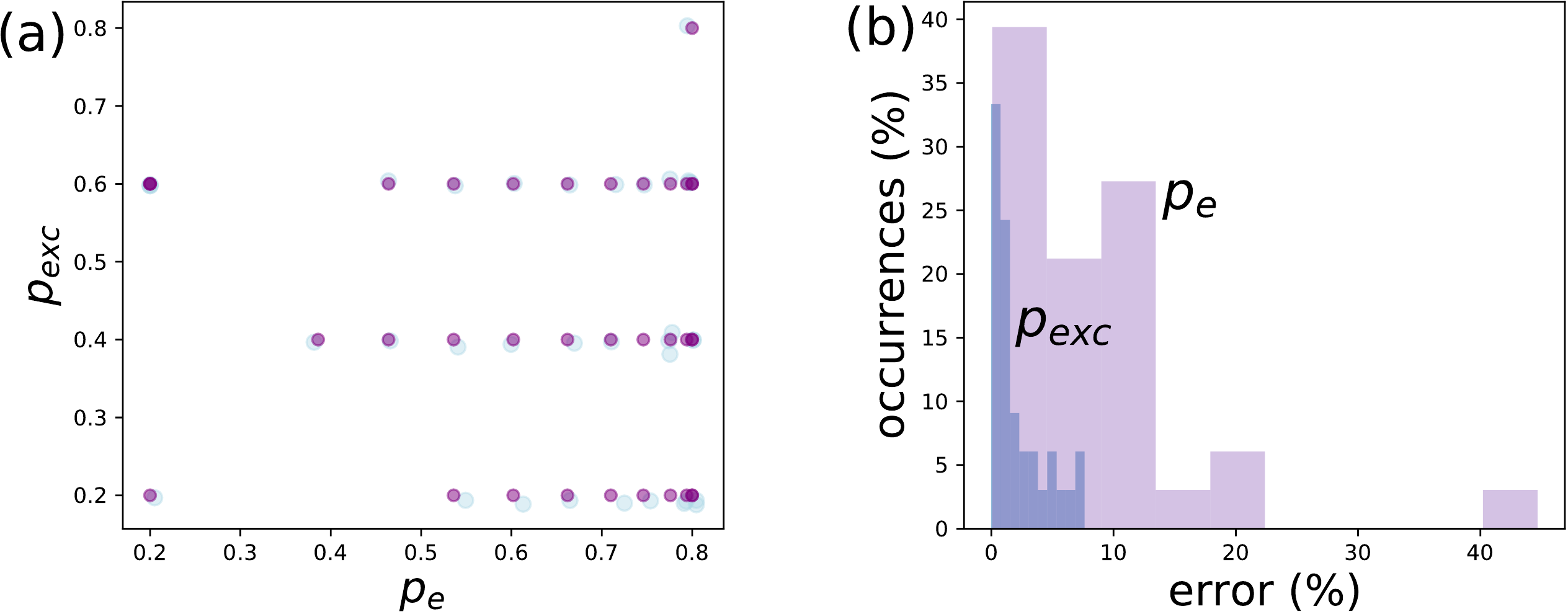}
\caption{Linearly polarized and right circularly polarized light
(a) predictions and (b) error distribution of
the experimental testing dataset for the training of the
sub-neural network comprised of layers 2-5.
}
\label{figure10}
\end{figure}

\begin{figure}
\flushright
\includegraphics[width=0.65\textwidth]{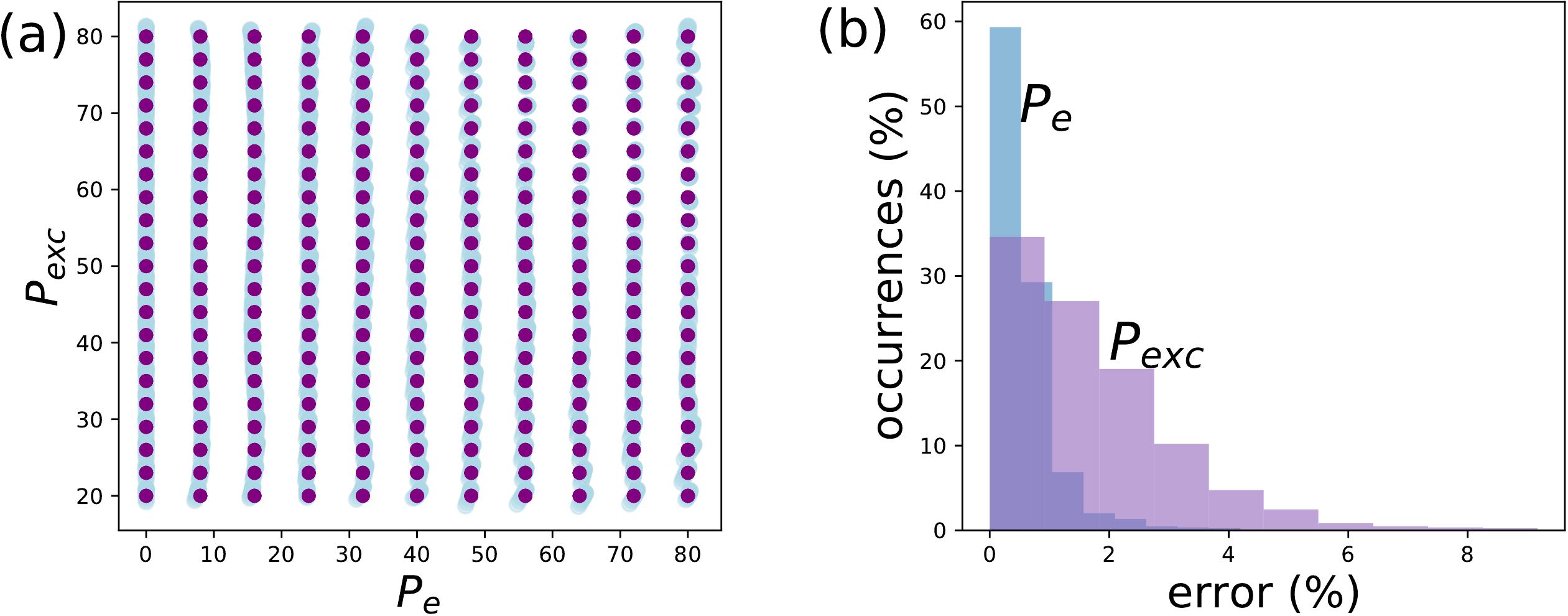}
\caption{Linearly polarized and right circularly polarized light
(a) predictions and (b) error distribution of
the theoretical testing dataset for the training of the
whole neural network comprised of layers 1-6.
}
\label{figure11}
\end{figure}

\begin{figure}
\flushright
\includegraphics[width=0.65\textwidth]{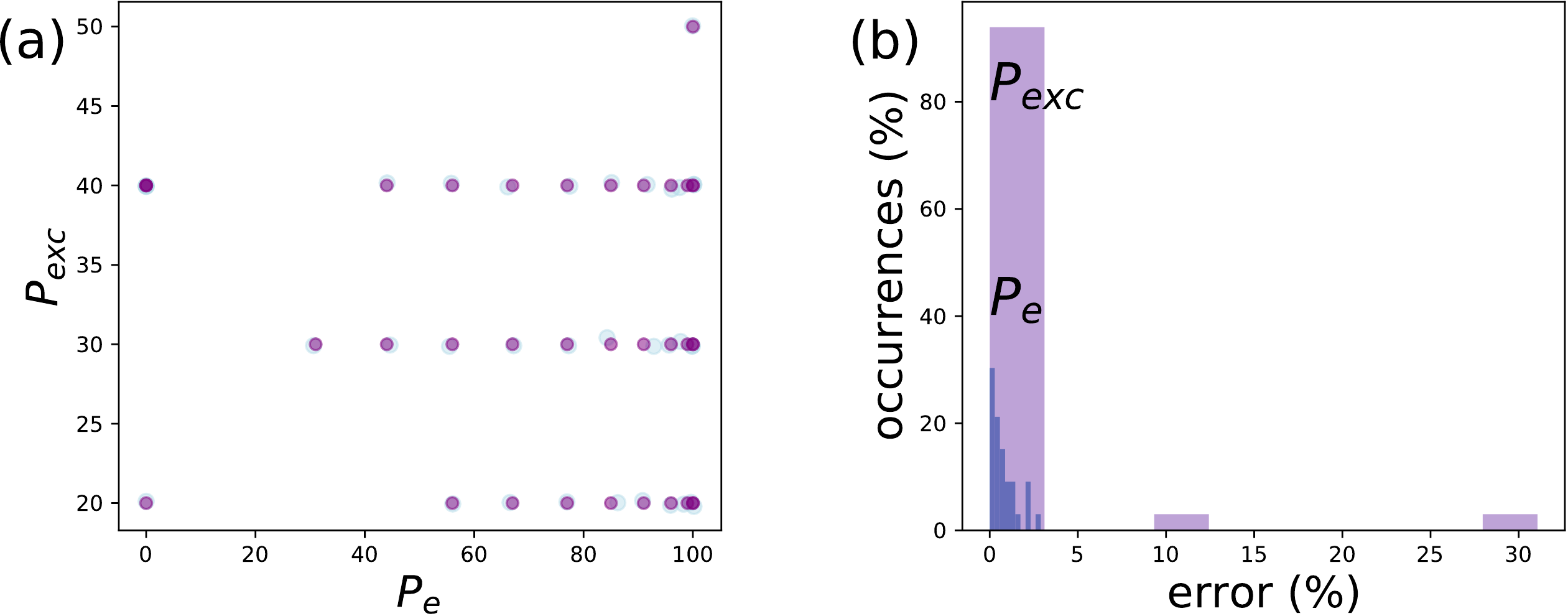}
\caption{Linearly polarized and right circularly polarized light
(a) predictions and (b) error distribution of
the experimental dataset for the training of the
whole neural network comprised of layers 1-6.
}
\label{figure12}
\end{figure}

Two functions are computed within each neuron.
First, the preactivation value of the $i$-th neuron in the
$k$-th layer is calculated as
$\sum_{j}^{k}w^{k-1}_{i,j}a^{k-1}_j+b^{k-1}_i$
where $w^{k}_{i,j}$ and $b^{k}_i$ are termed weights.
Then, the postactivation value is obtained
by applying the activation function $\Phi$
to the preactivation value.
In this way,
\begin{eqnarray}
a^k_i &=& \Phi_{\mathrm{ReLU}}
  \left(\sum_{j=1}^{n_k} w^{k}_{i,j}a^{k-1}_j+b^1_i\right),\,
  k=1,5, \\
a^k_i &=& \Phi_{\mathrm{sigmoid}}
  \left(\sum_{j=1}^{n_k} w^{k}_{i,j}a^{k-1}_j+b^2_i\right),\,
  k=2,3,4,
\end{eqnarray}
where $a^0_i=\Gamma_i$, $a^1_i=\gamma_i$,
$a^4_i=(p_e,p_{\mathrm{exc}})$,
$a^5_i=(P_e,P_{\mathrm{exc}})$ and $a^{2,3}_{i}$
are the activation values of the auxiliary neurons
of the hidden layers. The numbers of neurons in
each layer are $n_1=n_2=3$, $n_3=12$, $n_4=12$, $n_5=n_6=2$.
The {\it ReLU} (Rectified Linear Unit) and {\it sigmoid} activation functions are given by
\begin{eqnarray}
    \Phi_{\mathrm{ReLU}}(x) &=& \max(0,x),\\
    \Phi_{\mathrm{sigmoid}}(x) &=& \frac{1}{1+\exp(-x)}.
\end{eqnarray}
By means of the {\it ReLU} activation function
the outer layers of the NN rescale
the input and output parameters.
The {\it sigmoid} activation function of the inner
layers helps to incorporate the nonlinear behaviour
of the intensity and degree of circular polarization.

The training process of the NN is not
as straight forward as the logistic regression's.
The sub-NN comprised
of layers 2-5 is initially trained using the normalized inputs
($\gamma_1$, $\gamma_2$ and $\gamma_3$)
and outputs ($p_{\mathrm{exc}}$ and $p_e$).
The normalization consists of a rescaling
of the input ($\Gamma_1$, $\Gamma_2$ and $\Gamma_3$)
and output ($P_{\mathrm{exc}}$ and $P_e$) parameters
in the range $[0.2,0.8]$. In this range
the {\it sigmoid} activation function behaves almost linearly,
avoiding the saturation regions $[0,0.2)$ and $(0.8,1]$.
In this manner, the normalized parameters take the form
\begin{eqnarray}
\gamma_i &=& 0.2
  +\frac{0.6}
  {\Gamma_{i}^{\mathrm{max}}-\Gamma_{i}^{\mathrm{min}}}
  \left(\Gamma_i-\Gamma_{i}^{\mathrm{min}}\right)\label{eq:renormfirst}\\
  p_{\mathrm{exp}} &=& 0.2
  +\frac{0.6}
  {P_{\mathrm{exp}}^{\mathrm{max}}-P_{\mathrm{exp}}^{\mathrm{min}}}
  \left(P_{\mathrm{exp}}-P_{\mathrm{exp}}^{\mathrm{min}}\right)\\
 p_e &=& 0.2
  +\frac{0.6}
  {P_e^{\mathrm{max}}-P_e^{\mathrm{min}}}
  \left(P_e-P_e^{\mathrm{min}}\right).
  \label{eq:renormlast}
\end{eqnarray}
The sub-NN parameters
are determined through the optimization of
the mean squared error loss function.
In this first phase of the training process
the initial weights are set randomly
as can be seen in the algorithm
provided in the supplementary material.

Figure \ref{figure9} (a) shows a sample
of the theoretical training points (purple dots) compared to the
estimated ones (light blue dots) for
right circularly polarized light and linearly polarized light.
The error distributions of
the rescaled outputs
$p_e$ (light blue)
and $p_{\mathrm{exp}}$ (purple)
from the testing
dataset are displayed in Fig. \ref{figure9} (b).
In this plot, the error is calculated
as the deviation of the estimated $p_e$
and $p_{\mathrm{exp}}$ with respect to the
true values provided by the testing datasets.
Similarly, Fig. \ref{figure10}
shows the predictions and the error distribution
produced by the experimental training
and testing datasets.

\begin{figure}
\flushright
\includegraphics[width=0.95\textwidth]{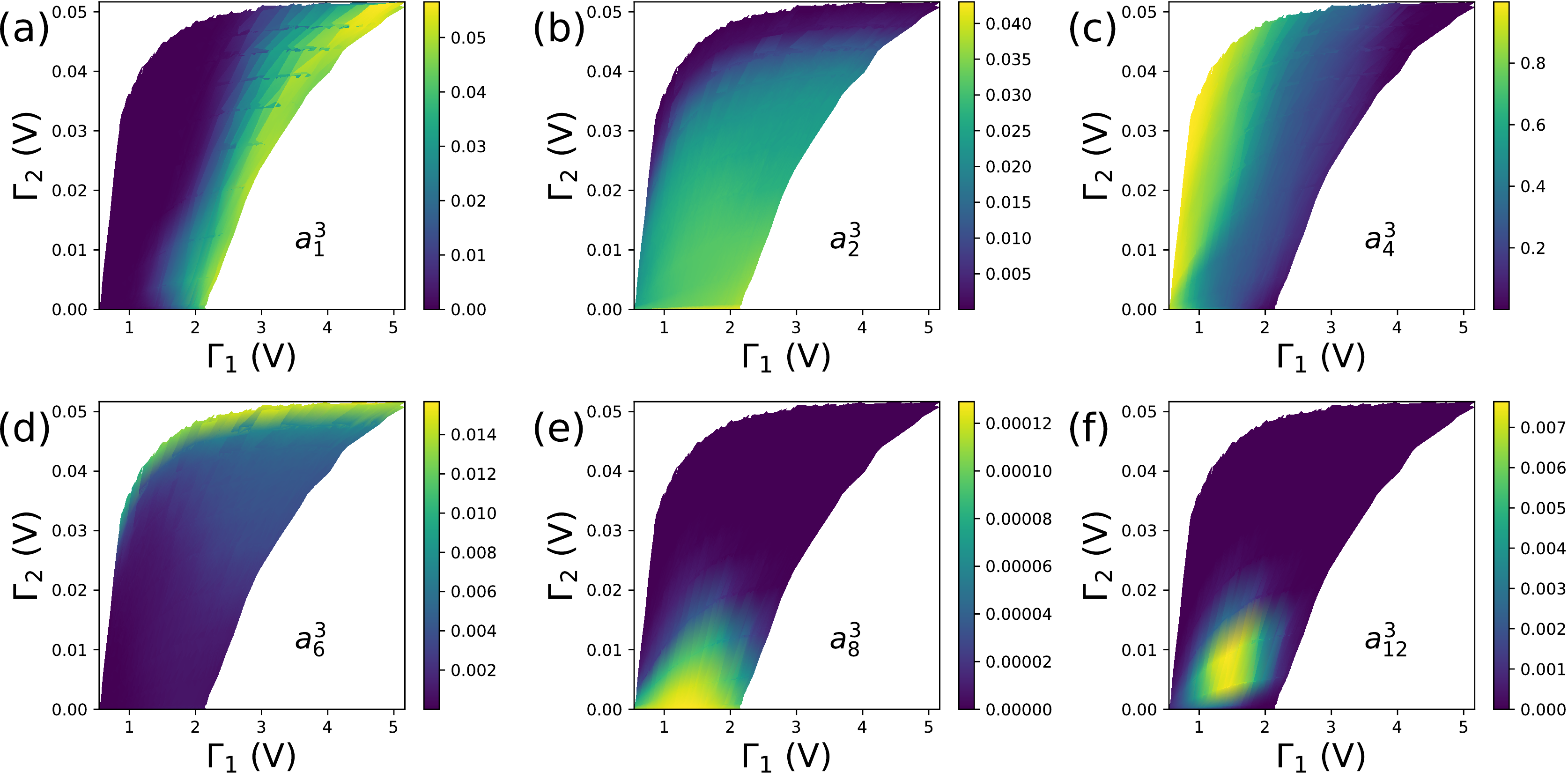}
\caption{Density plot of the activation values $a_n^3$ in
the $\Gamma_1$, $\Gamma_2$
parameter space of some of the neurons
in the fourth layer. The activations
(a) $a^3_1$, (b) $a^3_2$, (c) $a^3_4$, (d) $a^3_6$ , (e) $a^3_8$ and (f) $a^3_{12}$
are shown.}
\label{figure13}
\end{figure}

In the next phase of the NN
training process the
model parameters are further optimized through the
outer {\it ReLU} layers.
The weights
$w^{2}_{i,j}$,
$w^{3}_{i,j}$, $w^{4}_{i,j}$, $b^{2}_{i}$,  $b^{3}_{i}$
and $b^{4}_{i}$
that result from the optimization
of the inner layers serve as seed
values for this training phase.
Likewise the initial values of the two outer layers
can be estimated from
Eqs. (\ref{eq:renormfirst})-(\ref{eq:renormlast}) as
\begin{eqnarray}
w^{1}_{i,j} &=& \delta_{i,j}\frac{0.6}
  {\Gamma_{i}^{\mathrm{max}}-\Gamma_{i}^{\mathrm{min}}},\\
b^{1}_{i} &=& 0.2-\frac{0.6}
  {\Gamma_{i}^{\mathrm{max}}-\Gamma_{i}^{\mathrm{min}}}
  \Gamma_{i}^{\mathrm{min}},\\
w^5_{1,1} &=& \frac{0.6}
  {P_e^{\mathrm{max}}-P_e^{\mathrm{min}}},\\
b^5_1 &=& 0.2 -\frac{0.6}
  {P_e^{\mathrm{max}}-P_e^{\mathrm{min}}}
  P_e^{\mathrm{min}},\\
w^5_{2,2} &=& \frac{0.6}
  {P_{\mathrm{exp}}^{\mathrm{max}}-P_{\mathrm{exp}}^{\mathrm{min}}},\\
b^5_2 &=& 0.2 
-\frac{0.6}
  {P_{\mathrm{exp}}^{\mathrm{max}}-P_{\mathrm{exp}}^{\mathrm{min}}}
  P_{\mathrm{exp}}^{\mathrm{min}}\\
w^5_{1,2} &=& w^5_{2,1} =0
\end{eqnarray}
The whole NN (layers 1-6) is optimized
through the mean squared error loss function
using the above mentioned
initial parameters. This further improves
the error distribution as can be seen
in Figs. \ref{figure11} and \ref{figure12} for the
theoretical and experimental datasets, respectively.
Note that
in the previous training phase
the error distributions are both sharpened
and displaced to the small error region
after the optimization of the
whole NN.
The theoretical dataset yields
$0.23\%$ and $2.5\%$
average prediction errors for
$P_e$ and $P_{\mathrm{exc}}$,
respectively.
As a result of the small number of instances,
the experimental dataset
gives larger prediction errors of
$1\%$ and $10\%$
although, as the error distribution
in Fig. \ref{figure12}
shows, $92\%$ (32 out of 34 instances) of these $P_{\mathrm{exc}}$ errors
are below $3\%$.

A very similar behaviour is observed for
left circularly polarized light.
The predictions and error distributions for this case
are not shown here but can be
seen in the algorithm presented in
the supplementary material.

Even though it is not essential for the
proper functioning
of the NN, it is interesting
to zoom in on the activation of each
of the neurons belonging to the hidden layer.
The activation $a_n^3$ of some of the neurons
in the hidden layer (layer 4) 
as a function of $\Gamma_1$ and $\Gamma_2$ can
be viewed in Fig. \ref{figure13}.
It can be observed that each neuron specializes in a
particular feature. For example, neurons $n=1$ and $n=4$
are very sensitive
to the high and low power isolines respectively.
Neurons $n=2$ and $n=6$ are, instead, responsive to the low and high
$P_e$ isolines. Other neurons, as $n=8$ and $n=12$ are rather
concentrated in very specific areas of the $\Gamma_1 - \Gamma_2$ parameter
space.

\section{Conclusions}
We have proposed and demonstrated through theoretical
and experimental data a machine learning algorithm
that converts the electrical input signals of a 
GaAs$_{1-x}$N$_x$ based
circular polarimeter
into intensity and degree of circular polarization.
The three input parameters might come
from the electrical signals of the PC
(voltage or current) or the PL
of the sample subject to three different
external magnetic fields ($-50$\,mT, $50$\,mT and $150$\,mT).
We have demonstrated that the logistic regression
can be used very effectively as a predictive model
for the handedness of light misclassifying less than
$1.5\%$ of the experimental instances and about
$0.1\%$ of the theoretical ones.
We have also proven that the six layer NN model
presented here
can be trained to predict the intensity and circular
degree of polarization of an incident beam of light
from the electrical input signals.
The NN is capable of discriminating
intensity and degree of circular polarization
within a range of incidence angles up to $25^\circ$.
The great majority ($92\%$) of the predictions made by the NN model
on the experimental dataset have less than $3\%$ error.
It has been shown that the functionality of other
polarimeter configurations can be significantly enhanced
by increasing the number of parameters
introduced into the machine learning model\cite{doi:10.1021/acsphotonics.8b00295}.
Therefore, it would be reasonable to
expect that the performance of the GaAs$_{1-x}$N$_x$
circular polarimeter could be further
improved by adding new parameters, for example,
the values of the PC at
other magnetic field intensities.
Even though the proposed method was analyzed
for the particular case of a 
 GaAs$_{1-x}$N$_x$
circular polarimeter we believe that
a generalization of this algorithm (for different numbers
of input or output parameters) could
improve the performance of a broad variety of optical sensors.

\section*{Acknowledgements}
A.A-P., A.K. gratefully 
appreciates the financial support of Departamento de Ciencias B\'asicas 
UAM-A grant numbers 2232214 and 2232215.
X.M. also thanks Institut
Universitaire de France.
We are indebted to L.A. Bakaleinikov, E. L. Ivchenko and
V. K. Kalevich for the comments on our paper and
the careful reading of the manuscript.

\section*{Bibliography}

%\bibliographystyle{iopart-num.bst}
%\bibliography{mainbiblio}

\providecommand{\newblock}{}

\end{document}